# Extended Security Risks in IP Networks

Using Transit Network Nodes for Hijacking Traffic

Daniel Kharitonov
Stanford University
Palo Alto CA
dkh at stanford edu

Oscar Ibatullin
TraceVector Inc
San Jose CA
oscar at tracevector dot net

*Abstract*

**Exploitation techniques targeting intermediate (transit) network nodes in public and private networks have been theoretically known and empirically proven to work for quite some time. However, very little effort has been made to look into the network-specific risks of compromising the Internet infrastructure to this date. In this publication, we describe several methods of hijacking live network traffic following a successful attack on a router or switch. We demonstrate that modern network platforms are capable of targeted traffic replication and redirection for online and offline analysis and modification, which can be a threat far greater than loss of service or other risks frequently associated with such exploits.**

I. INTRODUCTION

After early days of the Internet predecessor ARPANET, transit network nodes were traditionally based on the heavily modified software packages, very similar to DEC RT-11 or various flavors of UNIX [1][2][3]. In today's network OS lineup, donor code from Linux, FreeBSD, VxWorks, and QNX still prevails [4] and is widely used for various internal and housekeeping errands such as TCP header compression, SNMP and telnet management, HTTP administrative tools, SSH access, and so on. As a result, the development of rootkits designed to gain unauthorized remote access for network nodes is not significantly different from that of general purpose computers and has been shown to be effective against a wide range of routers and switches—including recent examples of such exploits of Huawei [5] and Cisco [6] products.

However, traditionally public and private IP nodes have not been popular targets for attackers. The reason is that routers and switches present little value as "jump hosts" unless they connect together zones of different security levels. Network nodes are also of marginal interest as botnet members (relatively weak processing power and slow management interfaces) or host platforms for local data collection (no stored information). The main security concern about routers and switches has been the possibility of hackers wreaking havoc in large networks by deliberately crashing systems or introducing errors into the running protocols. For example, the National Institute of Standards and Technology (NIST) Common Vulnerabilities and Exposures (CVE) database [7] lists no less than 40 records for vulnerabilities allowing remote attackers to crash Cisco IOS and IOS XR devices; another 159 documents describe denial-of-service (DoS) techniques capable of exhausting critical resources. At the same time, there are only five known "administrative access class" exploits listed for IOS, mostly related to hardcoded or default SNMP and HTTP access procedures.[1] By comparison, NIST CVE queries for a general purpose OS (Windows NT) reveal 168 DoS exploits. Simultaneously, there are no less than 55 records of critical vulnerabilities on Windows NT potentially leading to unauthorized remote access. While this might be indicative of a wider set of attack opportunities inside a general purpose OS, there is also a strong chance that hackers are simply less interested in gaining access to the network infrastructure devices. This assumption is further supported by ready-made shellcode repositories. The Metasploit auxiliary module and exploit database, for example, contains only one Cisco IOS rootkit module compared to hundreds of modules developed for general-purpose host operating systems [8].

While the attacks on the routing and switching infrastructure may be harder to organize than consumer-grade exploitation of computers, they may offer an easier and quicker way to mount Advanced Persistent Threat (APT) assaults. Relative to "traditional" APT techniques, abuse of network infrastructure is much harder to track and discover with both host-level security software (virus scanners, job monitors, and so on) and Layer 7 network security devices such as intrusion detection systems (IDS).

Although transit data intercept can be used for mass scale attacks on consumers (for example, identity theft), its main threat may be in the area of industrial and financial espionage. For example, in a recent string of attacks known as "Operation Aurora," several companies (including Google and Adobe) reported loss of confidential information and intellectual property. Assuming the structure of this security incident is now properly understood, the corporate computers were infected through a zero-day vulnerability in a Web browser and downloaded malware specifically targeting source code repositories [9]. Such an attack could have been faster to stage and proliferate if intruders had the ability to inject content (for example, poisoned HTML frames) into user traffic regardless of computer location or connection endpoints, which would give attackers a bigger window of opportunity.

---

[1] We have selected Cisco IOS merely based on length of the database record; other network vendors might have similar trends.

At the same time, while the topic of traffic eavesdropping and intercept has been well studied for wireless networks [10][11][12], authors are not aware of recent studies that would demonstrate the feasibility of man-in-the-middle attacks against wide area network infrastructure. This gives an impression that manipulation of transit data in generic IP networks is too hard or even impossible without physical penetration.

## II. ROUTER AND SWITCH ARCHITECTURE

To understand why the remote manipulation of transit data is not a trivial task, it helps to look at the architecture of a modern router or switch with separation of control and forwarding planes (Figure 1). Such designs first became available in the late 1990s and quickly became the driving factor behind the unconstrained growth of the Internet. Instead of processing packets in the software forwarding path of earlier network devices, modern routers and switches employ network processors and ASICs to packet lookup and transfer functions, while reserving general purpose processors strictly for path computation, configuration, and management purposes.

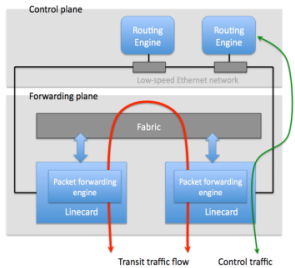

Figure 1. Control and forwarding plane separation in modern routers

In a modern router or switch as seen in Figure 1, transit traffic does not leave the hardware forwarding plane, where network processing units (NPUs) parse packet headers while leaving the rest of the payload intact (Figure 2). While it is possible to program NPUs to punt certain packets to a control plane (Routing Engine), most network devices are designed with security considerations in mind that allow only a fixed number of header bytes to reach the control plane. This limit allows for flow analysis and billing, but it excludes any possibility for payload eavesdropping, even after gaining full access to a router or switch.

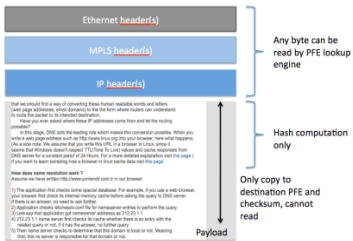

Figure 2. Packet structure and the lookup boundaries

Therefore, analysis and modification of transit traffic on a router or switch in principle requires processing outside of the forwarding plane. In the remainder of this paper, we will discuss methods and ways of achieving this task.

## III. AIDED PAYLOAD ATTACKS

Throughout this paper, we will assume that the first stage attack (node infiltration) is already complete and the attacker is able to execute arbitrary code on a target. In this section, we will detail an architecture that can be used to execute a second stage attack aiming to analyze and/or modify traffic flowing through the network device. Such actions can be variably used to gain access to data served by this node—hijacking active e-mail, social network, and remote access sessions; gathering personal and corporate Web access profiles and stats; injecting malicious code into running HTTP and database connections; and so on.

The most straightforward way to do this would be to utilize "service plane" hardware on the router or switch itself (Figure 3). Such hardware is used by OEM vendors to provide Layer 4 to Layer 7 packet services like deep packet inspection, stateful firewalls, and virus filtering, and in principle consists of high-performance CPU clusters capable of processing traffic at gigabit speeds [13].

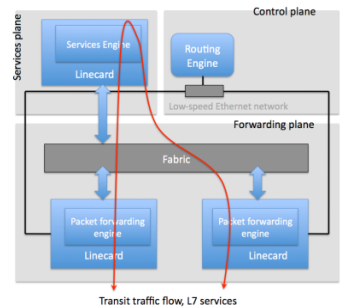

Figure 3. Payload traffic processing in the services plane

However, such hardware is relatively rare (found on less than one percent of network nodes[2]) and mostly proprietary, which makes onboard programming difficult for the attacker.

For this reason, we will exclude the services plane from further consideration in this paper and will concentrate on techniques that allow for traffic capture and processing aided by external computers. Such computers should be trivially available for attackers in the form of their own resources or (more likely) previously compromised hosts. However, as external traffic processors likely belong to the networks remote to the compromised network node, the main challenge is to deliver the "interesting" traffic to them in a manner that does not disrupt data flow and allows the eavesdropped connection to continue.

For such purpose, we principally distinguish two different techniques—*traffic redirection* and *traffic replication*.

---

[2] Based on shipping quantities estimated by Juniper Networks

## A. Traffic redirection

Traffic redirection exploits the fact that Internet connections are oblivious to the network path and can use asymmetric routing. This allows for a distributed man-in-the-middle attack using redirection. To create a redirected stream, the attacker must first identify the "interesting" traffic and funnel it into a captive filter applied to the ingress interface. Such a filter is normally designed to single out packets destined for a specific server (for instance, a popular resource vulnerable to HTTP session hijacking attack), specific protocol with little encryption (telnet or FTP), or specific source/destination addresses for scans (for example, potential victim's IP range).

Next, a feature called filter-based forwarding (FBF) or policy-based routing (PBR) is chained to the hardware filter entry to send the "interesting" traffic away from its normal destination based on the forwarding table and into a specially crafted IP-in-IP tunnel terminating at the aid host.

The aid (traffic processing computer) removes the outer IP header wrapper, scans (or stores) the payload, and sends packets back into the network with original source and destination IP addresses unmodified. Therefore, the intercepted traffic reaches its intended destination after a layover stop orchestrated by the attacker (Figure 4). This technique can be used on widely available routers and L3 switches from major vendors and only requires support for filter-based routing and IP tunneling such as generic routing encapsulation (GRE), both being the mainstay features.

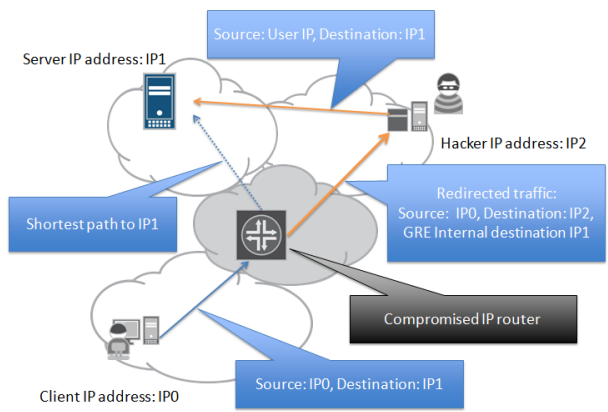

Figure 4.  Traffic redirection using IP tunnel

Moreover, as the captive filter can be changed at any moment, an attacker can quickly detail it based on initial analysis and thus avoid moving and scanning the large amounts of data. For example, if it is determined that the targeted user frequently accesses certain social and news sites, filter modifiers may be added to focus on Web traffic for specific addresses. Likewise, a rolling filter can be used to scan for traffic based on the source IP block assigned by a cable or DSL access provider. In that case, connections from access network tenants (families or individuals) can be singled out for examination and subsequent offline analysis, decryption, or further development of the ongoing APT attack.

The same redirection technique can be used in the return direction (server to user), where the entire protocol sessions (such as HTTP 1.0) can be hijacked and impregnated with viral code. Most obviously, this allows for large-scale zero-day attacks on the users previously identified as operators of vulnerable clients. This capability makes aided payload attacks through network infrastructure very efficient and very dangerous—instead of relying on drive-by techniques, an attacker can push viral content onto Internet users regardless of where they go as long as their traffic crosses the infected network node.

The relative ease of traffic redirection practically guarantees that this attack can be planned using any transit network device, and makes attack recognition challenging even when source and destination reside in trusted networks and behind strong firewalls.

However, traffic redirection also has some practical difficulties.

For one thing, if the remote aid host resides in (or behind) a network that supports source checking via filter or a reverse path forwarding (RPF) feature, this renders direct flow from IP2 to IP1 using IP0 as source impossible. If this happens, an attacker will have to establish an aid host in the same network where the source or destination of traffic resides. This is normally not a challenge when attacking access networks with multitudes of weakly protected computers, but it can be a challenge when dealing with traffic sourced from (and going to) corporate networks. In this latter case, an attacker may have to send the processed traffic back to the compromised router in the opposite direction of the IP tunnel and route it as usual from this point on. Such modification of topology would require a slightly more involved router programming.

For another example, the layover nature of traffic redirect attacks may cause bandwidth or delay artifacts. This may trigger dissent or complaints from affected users and compromise the attacker. Again, such concerns are more likely to surface when targeting corporate traffic, where it is common to see a significant amount of data within high-speed encrypted tunnels. To record such data for analysis without impacting timing of the flow, an attacker may have to resort to another technique called *traffic replication*.

## B. Traffic replication

When using replication, an attacker forks traffic instead of redirecting and only sends a copy to processing aids (Figure 5). This means delay but neither bandwidth nor network path attributes of the original flow change, making this method of attack completely invisible to affected connections. Since original packets are allowed to proceed to their destination, traffic forking is not capable of payload injection and remains strictly a method for eavesdropping.

It is relatively common for routers or switches to support native tunneling of replicated packets as a mechanism for lawful packet intercept [14]. In principle such capability can be reused for unauthorized packet forking, with a remote computer simulating a legitimate mediation host and traffic analyzer. However, practically speaking, lawful intercept is not very useful for attackers in its native form as it involves significant modification of a router configuration and use of

IPsec for delivery of replicated traffic. Therefore, attackers will likely resort to less complex mechanisms, such as using packet replication and GRE tunnels.

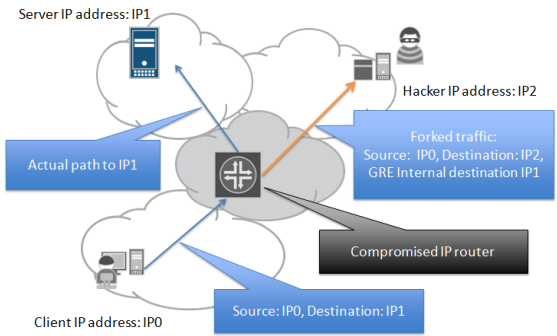

Figure 5.  Traffic replication (principle scheme)

Most of today's routers and switches are capable of making packet copies, which under normal circumstances belong to either L3 multicast trees such as IP and MPLS distribution paths, or L2 broadcast flooding domains such as bridged Ethernet network or virtual private LAN service (VPLS) mesh. To make packet forking work, an attacker may employ either mechanism, masquerading it as legitimate forwarding entries. However, L3 replication normally involves creation of forwarding states network-wide, such as point-to-multipoint label–switched paths (P2MP LSPs), which may be impractical or too risky for an attacker. A much simpler L2 replication scheme can be employed at the peering points, where multiple routers are interconnected with Ethernet switches. If an attacker controls routers R1, R2, and R3, an FBF entry for incoming packets from IP0 to IP1 can be matched to a next hop toward S1 with hardcoded multicast media access control (MAC) as a destination address. This will force S1 to replicate the packet on all ports (a normal behavior for unknown multicast groups). Routers R2 and R3 will both receive the packet; R2 will forward it as usual; and R3 will send it into a tunnel towards IP2 via FBF entry.

To achieve this, it is sufficient to program R2 and R3 Ethernet interfaces towards S1 for multicast receiver MAC address matching of the L2 next-hop entry on R1. This should be enough for packets on R2 and R3 to clear MAC ternary content addressable memory (TCAM) and be sent to the L3 lookup engine. If there are other routers connected to the same S1 switch, they will ignore their packet replicas due to unknown multicast destination.

Note that switch S1 does not necessarily have to be in the upstream direction towards IP1. If R2 receives the packet and determines it must go via another router (for example, R1), it will simply send it back. Therefore, destination IP1 is guaranteed to receive only one copy of the flow.

Although this scheme involves significantly more equipment and requires access to an L2 broadcast domain, it also avoids tampering with the control plane of the entire network, as all necessary changes can be programmed locally on routers R1, R2, and R3. Switch S1 needs no special configuration and does not have to be compromised to achieve the result.

## IV. PRACTICAL IMPLEMENTATION DETAILS

So far we have discussed the principles and architecture of attacks on traffic payloads from compromised network nodes. One practical consideration when designing such an attack would be the ability to conceal the ongoing activity and allow the network node to function as usual. This means that the most straightforward method of exploitation—direct reconfiguration of a router or switch—is also the less preferred as configuration management is normally a closely monitored area of operation.

Therefore, an attacker will likely aim to leave no trace in configuration files and inject the forwarding state changes as runtime (ephemeral) entries with no human readable form. This method is quite robust with respect to persistency because modern network nodes are not rebooted or reconfigured frequently and can accumulate uptime measured in months or years of continuous operation. With unified in-service software upgrade (unified ISSU) in place, routers and switches can even pick software fixes without resetting the running forwarding table databases.

To program ephemeral forwarding entries for packet redirection or packet replication, one needs to be able to install filters and next hops to the forwarding tables of a running system. In principle, this involves sending a crafted command or communication signal to a process that handles the required function. Since the internal details and APIs of such communications are not public, the attacker will have to gain access to vendors' software development kit (SDK) and third-party development toolkits (if available) or intercept and reverse engineer communication pipes to figure out the commands and arguments needed for ephemeral state changes. This task is much less complex than it sounds, because the initial analysis is easy to do on a test system by simply tracking the logs and messages of processes handling a routine PBR or FBF configuration change. In addition, on many routers and switches, the forwarding plane itself runs a low-level operating system (line-card kernel) where a rich set of command-line interface (CLI) commands is available to verify and manipulate entries in the forwarding table directly (Figure 7).

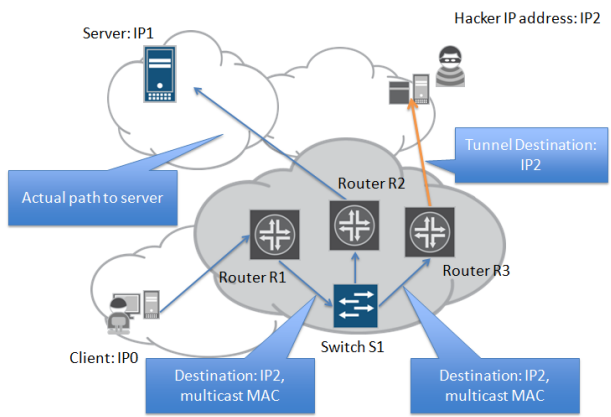

Figure 6.  L2-based replication using two routers and a switch

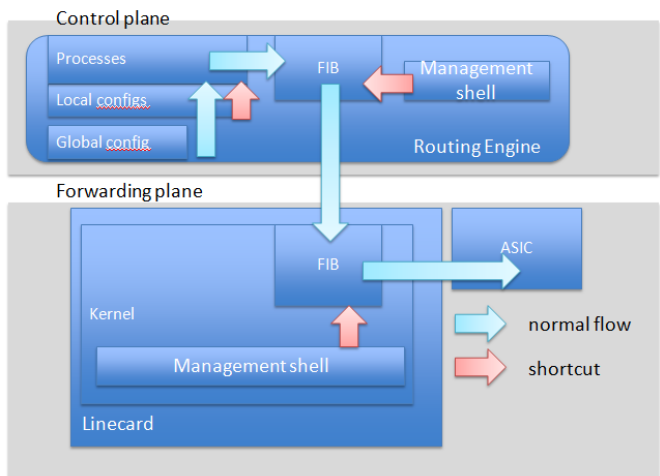

Figure 7. Injecting ephemeral states into control and forwarding planes

In fact, in a typical network operating system, an attacker has the choice between locating and modifying local management configuration databases for system processes, direct modification of main forwarding table entries from a management shell, and direct modification of local forwarding table entries from the management shell of a line card. This diversity of points for tampering gives an attacker a high chance of success.

Once a working combination of instructions is found in the physical lab or using a vendor SDK in a virtual/simulation environment, it can be easily transferred to the running systems by the same vendor.

## V. Aided Payload Attacks and SDN

The software defined network (SDN) is an emerging approach to control plane programming that allows for centralization of (previously distributed) control plane path computation. While there is no single definition of SDN architecture, it is most often described as the ability to program forwarding planes of the network nodes directly from external control servers and appliances, for an increased degree of control and flexibility.

While the control channel of the SDN connection to the device may (or may not) be an attractive target for a first stage attack on a network node, in this paper we are primarily interested in the fact that underlying SDN protocols such as Path Computation Element (PCE) [16] and OpenFlow use APIs for creating ephemeral forwarding states. In particular, Openflow version 1.3 defines functionality of policy routing and tunnel (IP-in-IP) connections [15]. Although SDN capabilities are not widely supported today, in the future we can reasonably expect this feature set to grow larger and cover the full range of functions needed for programming of payload attacks.

Existence of SDN connectors capable of forming ephemeral states on a router or switch makes the attacker's life easier when creating traffic forks and redirects, because it allows exploit toolkits to be less specific to the software architecture of the target. For example, instead of decrypting proprietary instructions to establish a captive filter for interesting traffic, an attacker can send a standardized command into an OpenFlow connector to achieve the same result.

Moreover, if the control server is compromised, it makes it possible to attack payloads handled by multitudes of controlled devices, effectively monitoring or modifying traffic within the entire network span. Therefore, proliferation of software-defined networks may add additional factors to the risks we have already described.

## Conclusion and Recommendations

In this article, we detail the description of a specific cyberattack type that was not previously observed on a large scale. We demonstrate that such an attack can be mounted against conventional routers and switches and does not require extraordinary knowledge of networks or specialized programming skills.

Therefore, in addition to standard security measures recommended for all network nodes (such as limiting administrative access by hardware-based filters applied to loopback interfaces, removal of unsafe and default access means and accounts, etc.), we can identify the security measures that can be employed to make redirection and replication more difficult

Use of open-source software (OSS) components in network equipment operating systems traditionally presents a potential vector for the first stage attack. While timely patching remains a very effective protection from this threat, it's often overlooked that the networking industry's release cycle cadence is relatively slow. Therefore, specialized network devices such as intrusion prevention systems (IPS) are recommended to protect against first stage attacks, especially where the value of transit data is high. To mitigate the cases where an attacker has succeeded in the first stage and has full access to the compromised network node, it is important to keep the node isolated so that the second stage has little effect on the rest of the infrastructure and data flow. As global Internet policy, we strongly recommend dropping traffic with wrong source addresses originating from leaf networks. For transit networks, loose RPF policies are always recommended on peering routers. This will force attackers to return traffic to the compromised node for re-injection into the network and will give security specialists more indicators of the ongoing activity.

For node-specific security enhancement, we would like to raise the importance of monitoring the ephemeral forwarding states, especially if they are unusual or unexpected for a node type such as FBF entries or IP-in-IP tunnels. Vendors should provide suitable commands and mechanisms for listing forwarding states by policy or next-hop class.

Finally, we would suggest that security officers and specialists pay special attention to network paths taken by critical connections such as branch VPN and data center traffic crossing public networks. Although the complete view of the path in the tunnels is partially hidden from end users, it is quite possible to discover unexpected intermediate nodes in the return leg of a redirected connection by using regular probing tools such as traceroute over TCP. If traceroute shows different results when working over TCP and Internet Control Message Protocol (ICMP), this might be a sign of divergent routing and a good reason for investigation.

Finally, we recommend elevated security measures when deploying SDN controllers. While most of controller implementations are based on PC hardware, they are frequently deployed alongside other servers or even end-user computers and thus become easy victims for virus and botnet outbreaks. Considering the potential damage arising from compromising SDN controllers, they need to reside on dedicated network segments denied of both Internet and Intranet access.